\documentclass{aa}
\usepackage{graphicx}
\begin{document}
   \title{Stellar evolution with rotation and magnetic fields:}

   \subtitle{III: The interplay of circulation and dynamo}

\author{Andr\'e Maeder, Georges Meynet}

     \institute{Geneva Observatory CH--1290 Sauverny, Switzerland\\
              email:  andre.maeder@obs.unige.ch\\
             email: georges.meynet@obs.unige.ch
               }

   \date{Received  / Accepted }

   \offprints{Andr\'e Maeder} 

   \abstract{We examine  the effects of the magnetic field created by the Tayler--Spruit dynamo
   in differentially rotating stars. Magnetic fields of the order of a few $10^4$ G are present 
   through most of the stellar envelope, with the exception of the outer layers.
   The diffusion coefficient for the transport of angular momentum is 
   very large and it imposes nearly solid body rotation during the MS phase. In turn, solid body rotation
    drives meridional circulation currents which are much faster than usual and leads to much larger 
    diffusion coefficients than the magnetic diffusivity for the chemical species. 
   The consequence is that the interplay of the thermal and magnetic instabilities
   favours the chemical transport of elements, while there would be no transport in models with magnetic field only.  We also discuss the effects on the stellar interior, lifetimes and HR diagram.

   }
 \titlerunning{Rotation and Magnetic Fields}              
   \maketitle
%

\section{Introduction}

Stellar rotation influences all the outputs of stellar evolution and nucleosynthesis and several grids of models have been made for massive stars
(Langer et al. \cite{langeragb}; Heger et al. \cite{hegerI}; Meynet \& Maeder \cite{MMV}).
The effects of rotation are even more important at metallicities $Z$ lower than solar (Maeder \& Meynet \cite {MMVII}).
 However, we do not know the  role of the magnetic field in stellar evolution 
 and it is the purpose of this series of works to approach this 
 matter, which is considered to be   critical one (cf. Roxburgh \cite{Rox03}). Spruit
 (\cite{Spruit02})  proposed a dynamo mechanism operating in 
 stellar radiative layers in differential rotation. This dynamo is based on 
 the conjectured Tayler instability, which is apparently the first  one to occur in a radiative zone (Tayler \cite{Tay73}; Pitts \& Tayler \cite{Pitts86}). For now, there is  no empirical or 
 observational proof of the existence of this instability. 
 According to Pitts \& Tayler (\cite{Pitts86}) and Spruit (\cite{Spruit02}),  even a very weak horizontal 
 magnetic field is subject to this instability, which then creates a vertical field component,
 which is wound up by differential rotation. As a result, the field lines become progressively
 closer and denser and thus a strong horizontal field is created at the
 energy expense of differential rotation.
 
 In a first paper (Maeder \& Meynet \cite{Magn1}, paper I), we have shown 
 that in a rotating star a magnetic field can be created during MS evolution 
 by the Spruit dynamo. We have examined the timescale for the field 
  creation, its amplitude and the  related diffusion coefficients. The clear result is that magnetic
  field and its effects are quite important. In the second paper (Maeder
  \& Meynet \cite{Magn2},  paper II), a generalisation of the  equations of the 
dynamo was developed. The solutions  fully agree with Spruit's solution in the two limiting
  cases  considered (Spruit  \cite{Spruit02}), i.e. 
  ``Case 0'' when the $\mu$--gradient
  dominates and ``Case 1'' when the $T$--gradient dominates with large non--adiabatic effects.
  Our more general solution encompasses all cases of $\mu$-- and $T$--gradients, 
  as well as all cases from the fully adiabatic to  non--adiabatic solutions. 
  
   Paper II  suggested that there is a complex feedback between the magnetic instability,
	which generates the field, and the thermal instability which drives the 	meridional circulation
	(Maeder \& Meynet \cite{Magn2}). However, it was beyond the scope of paper II to make numerical models 
	of  the interaction between  circulation currents and  dynamo. In this paper we  account for this feedback, the  main steps of which are the following:
	\begin{itemize}
 \item	Differential rotation creates the magnetic  field.
 \item  Magnetic field tends to suppress  differential rotation.
 \item  The absence of differential rotation  strongly  enhances meridional circulation,
 \item  Meridional circulation  increases differential rotation.
 \item  Differential rotation creates the magnetic field (the loop is closed).
 \end{itemize}
 
  Thus, a delicate balance between the thermal and magnetic effects occurs   during evolution.
  The balance of these effects influences the transport of chemical elements to the stellar surface 
  and the internal transport of angular momentum, which determines the evolution of the surface rotation.

 In Sect. 2, we collect in a  consistent way the basic equations of the dynamo.
 In Sect. 3, we give the basic expressions for the transport coefficients.  
  In Sect. 4, we calculate numerical models for the interaction of the dynamo and circulation on the internal stellar structure.
  In Sect. 5, the effects at the stellar surface and in particular the evolution of abundances are analysed.
  Sect. 6  gives the conclusions.

\section{The general equations for the dynamo}  \label{dyn}

The present set of equations for the dynamo based on Tayler--Spruit  instability has three important 
advantages with respect to the system of equations  by Spruit (\cite{Spruit02}).
\begin{enumerate}
	\item The equations by Spruit apply to the non--adiabatic case. As most of the stellar interior is adiabatic, this situation is not appropriate. The present equations fully treat the problem whether it is adiabatic or non--adiabatic.
	\item Either the $\mu$--gradient was considered to dominate or it was absent. Such 
	a situation is not  adequate  when mild mixing effects, which produce small 
	$\mu$--gradients, are studied. In addition, as always in all problems of mixing, the $\mu$--gradients need to be 
	very carefully  treated, otherwise the results are incorrect.
	\item There is no need to interpolate between the two asymptotic solutions called ``Case 0'' or ``Case 1''. The interpolation was made by Spruit (\cite{Spruit02}) as a function of the parameter $q$, which defines the amount of differential rotation (see definition given in Eq.~\ref{qo}).  This is hazardous, as it expresses the  interpolation
	of effects concerning  the $\mu$--gradient and the degree of non--adiabaticity as a function of 
	differential rotation. This is not physically justified and it introduces an additional spurious 
	dependence on the parameter $q$ into the problem.
\end{enumerate}

\noindent
Let us briefly summarize the consistent system of equations.
 The energy density $u_{\mathrm{B}}$ of a magnetic field of intensity $B$ per volume unity is
 
  \begin{eqnarray}
u_{\mathrm{B}} = \frac{B^2}{8  \; \pi} \,  = \,{1 \over 2} \; \rho \; r^2 \omega_{\mathrm{A}}^2 \,
\quad \mathrm{with} \quad	\omega_{\mathrm{A}}= \frac{B}{(4\; \pi \rho)^{1\over 2} \;r}  \;	,
 \label{def1}
\end{eqnarray}

  \noindent
where $\omega_{\mathrm{A}}$ is  the  Alfv\'en frequency   in a spherical geometry, $r$ is the radius and
$\rho$ the density.
In stable radiative layers, there is in principle no particular motion. However, if due to the
magnetic field or rotation, some  unstable displacements
 of vertical amplitude $l/2$ occur around an average stable position,
 the restoring buoyancy force produces  vertical oscillations around the equilibrium position with a
frequency equal to the Brunt--V\H{a}is\H{a}l\H{a} frequency $N$.  In a medium 
 with both thermal and magnetic diffusivity $K$ and $\eta$, 
  the Brunt--V\"{a}is\"{a}l\"{a} frequency becomes (Maeder \& Meynet \cite{Magn2}) 
  
\begin{equation}
N^2 = \frac{\eta / K} {\eta / K  +2} \; N^2_{\mathrm{T}}+ N^2_{\mu}  \; ,
\label{Nfinal}
\end{equation}

\noindent
 with the radiative diffusivity  $K= \frac{4ac T^3}{3 \kappa \rho^2 C_{\mathrm{p}}}$ and 
 with the following partial frequencies 
\begin{equation}
N_{\mu}^2 \; = \; \frac{g \varphi}{H_{\mathrm{P}}} \nabla_{\mu}  \quad \mathrm{and} \quad
N_{\mathrm{T}}^{2}\; = \; \frac{g \delta}{H_{\mathrm{P}}}(\nabla_{\mathrm{ad}}-\nabla) \; .
\label{NmuT}
\end{equation}

\noindent
The thermodynamic coefficients $\delta$ and $\varphi$ are 
$\delta= - \left(\frac{\partial \ln \rho}{\partial \ln T}\right)_{\mathrm{P},\mu}$ and
$\varphi=  \left(\frac{\partial \ln \rho}{\partial \ln \mu}\right)_{\mathrm{T,P}}$. $H_{\mathrm{P}}$ is the pressure scale height.
The restoring oscillations will have an average density of kinetic energy 

\begin{equation}
u_{\mathrm{N}}  \simeq  \; f_{\mathrm{N}} \; \rho \; l^2 \; N^2  \; ,
\label{uN}
\end{equation}

\noindent
where $f_{\mathrm{N}}$ is  a geometrical factor of the order of unity.
If the magnetic field produces some instability with a vertical component, one must have
 $u_{\mathrm{B}} >  u_{\mathrm{N}}$.
 Otherwise, the restoring force of gravity which acts at the dynamical timescale
 would immediately counteract the magnetic instability. From this inequality, one obtains 
$ l^2 < \frac{1}{2f_{\mathrm{N}}} \;r^2 \; \frac{\omega^2_{\mathrm{A}}}{N^2}$.
 If $f_{\mathrm{N}}= {1 \over 2}$, we  have the condition for the vertical amplitude of the  instability
 (Spruit \cite{Spruit02}; Eq.~6),
 
 \begin{equation}
 l <  \; r \; \frac{\omega_{\mathrm{A}}}{N}  \; ,
 \label{lr}
 \end{equation}
 
  \noindent
  where  $r$ is the radius.
  This means that there is a maximum size of the vertical length $l$ of a magnetic instability. In order not to be quickly  damped 
  by magnetic diffusivity, the vertical length scale of the instability must satisfy 
  
  \begin{equation}
 l^2 >  \frac{\eta}{\sigma_{\mathrm{B}}}= 
 \frac{\eta \; \Omega} {\omega_{\mathrm{A}}^2}  \; , 
 \label{lmin}
 \end{equation} 
 
 \noindent
 where $\Omega$ is the angular velocity and  
 $\sigma_{\mathrm{B}}$ is the characteristic frequency of the magnetic field.
 In a rotating star, this frequency is  $\sigma_{\mathrm{B}} =
 (\omega_{\mathrm{A}}^2 / \Omega)$  due to the  Coriolis force   (Spruit \cite{Spruit02}; see also Pitts \& Tayler \cite{Pitts86}). 
The combination of the limits given by Eqs. (\ref{lr}) and (\ref{lmin}) gives for the case of marginal stability,

\begin{equation}
\left(\frac{\omega_{\mathrm{A}}}{\Omega}\right)^4  =  \frac{N^2}{\Omega^2} \;
\frac{\eta}{r^2 \; \Omega}   \; .
\label{premier}
\end{equation} 

The equality of the amplification time of Tayler instability
  $\tau_{\mathrm{a}}=  N /(\omega_{\mathrm{A}} \Omega q)$  with {\bf{the inverse of the}} characteristic frequency 
  $\sigma_{\mathrm{B}}$ of the magnetic field leads to the   equation (Spruit \cite{Spruit02})
  
\begin{eqnarray}
\frac{\omega_{\mathrm{A}}}{\Omega} = q \; \frac{\Omega}{N}
  \quad
\mathrm{with} \quad q= -\frac{\partial \ln \Omega}{\partial \ln r}  \; .
\label{qo}
\end{eqnarray} 
\noindent
With account of the Brunt--V\"{a}is\"{a}l\"{a}  (Eq.~\ref{Nfinal}), one has 

\begin{equation}
\left(\frac{\omega_{\mathrm{A}}}{\Omega}\right)^2  =
\frac{\Omega^2 \; q^2}
{  N^2_{\mathrm{T}} \; \frac{\eta / K} {\eta / K  \;+ \; 2} + N^2_{\mu}}.
\label{deusse}
\end{equation} 

\noindent
By   eliminating  the expression of $N^2$ between Eqs. (\ref{premier}) and (\ref{deusse}), we obtain an expression for the magnetic diffusivity, 

\begin{equation}
\eta \;= \; \frac{r^2 \; \Omega}{q^2} \; \left( \frac{\omega_{\mathrm{A}}}
{\Omega}\right)^6  \; .
\label{eta}
\end{equation}

\noindent
Eqs. (\ref{premier})  and (\ref{deusse}) form a coupled system relating the two unknown 
quantities $\eta$ and $\omega_{\mathrm{A}}$. Instead, one may also consider for example the system
formed by Eqs. (\ref{deusse}) and (\ref{eta}).
Formally, if one accounts for the complete expressions of the thermal gradient $\nabla$, the system of equations would be
of degree 10 in the unknown quantity  $x=\left(\frac{\omega_{\mathrm{A}}}{\Omega}\right)^2$ (paper II).  
The fact that the ratio $\eta/K$ is very small allows us to bring these coupled equations to a system of  degree 4,

\begin{equation}
\frac{r^2 \Omega}{q^2 K} \left(N_{\mathrm{T}}^2 + N_{\mu}^2 \right)  x^4-
\frac{r^2 \Omega^3}{K} x^3 + 2 N_{\mu}^2 \; x - 2 \Omega^2 q^2 = 0 \; .
\label{equx}
\end{equation} 

\noindent
 The 
 solution of this equation, which is easily obtained numerically,  provides the 
 Alfv\'en frequency and by Eq. (\ref{eta}) the thermal diffusivity. As shown in paper II,  the various peculiar cases studied by Spruit (\cite{Spruit02}) are all contained in the more general solution given here.
 
   \begin{figure}[t]
  \resizebox{\hsize}{!}{\includegraphics{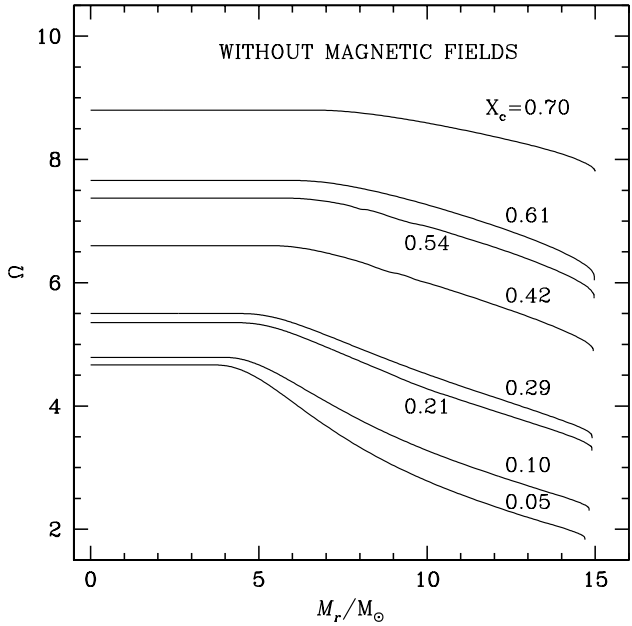}}
  \caption{Internal distribution of the angular velocity $\Omega(r)$ 
  as a function of the lagrangian mass in solar units in a 15 M$_\odot$ model, without magnetic fields, 
  at various stages of the model 
  evolution indicated by the central H--content $X_{\mathrm{c}}$ during the MS--phase. 
  The initial velocity $\upsilon_{\rm ini}$ = 300 km s$^{-1}$.}
\label{profoSH}
\end{figure}

 \begin{figure}[t]
  \resizebox{\hsize}{!}{\includegraphics{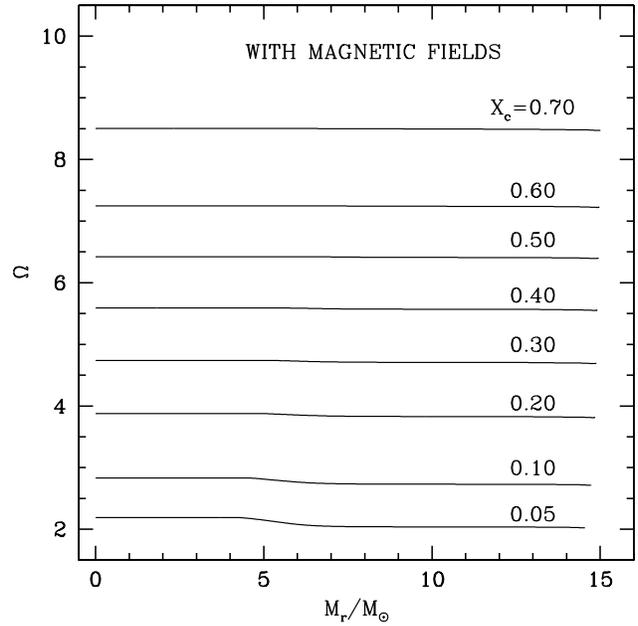}}
  \caption{Same as in Fig.~\ref{profoSH} but with magnetic fields. One notices the almost constant
  values of the angular velocity in the models.}
\label{profoH}
\end{figure}

\section{Coefficients for the transport of angular momentum and chemical elements}

 The condition that the ratio $\omega_{\mathrm{A}}/\Omega$ given by (Eq. \ref{qo}) is equal to or larger than 
 the minimum value defined by Eq. (\ref{premier}) leads to a condition on the minimum shear for the
 dynamo to work (cf. Spruit \cite{Spruit02})
 \begin{equation}
q > \left({N \over \Omega}\right)^{7/4} \left({\eta \over r^2 N}\right)^{1/4}\; .
\label{cond1}
\end{equation}
\noindent
There, $N$ is given by Eq. (\ref{Nfinal}) above and $\eta$ by Eq. (\ref{eta}). If this condition is not fulfilled, there is  no stationary situation and we consider that there is no magnetic field present.
In practice, we will see that this situation occurs in the outer stellar envelope.  
There is a second condition. We need to check that $\Omega > \omega_{\mathrm{A}}$.
If this last condition is not realized, the present system of equations does not apply and we should consider the case of very slow rotation. This case is treated in the Appendix. In the present models, the value of $\Omega$ considered
is large enough so that we did not have  to apply  the solution
for the case of very slow rotation.

The azimuthal  component of the magnetic field is much stronger that the radial one in the Tayler--Spruit dynamo. We have for these components (Spruit \cite{Spruit02})

 \begin{equation}
  B_{\varphi}= (4 \pi \rho)^{\frac{1}{2}} \; r \; \omega_{\mathrm{A}} \quad \mathrm{and} \quad
 B_ {\mathrm{r}}= B_{\varphi} \; (l_{\mathrm{r}} / r)  \; ,
 \label{champ}
 \end{equation}
 
 \noindent
 where $\omega_{\mathrm{A}}$ is the solution of the general
 equation (\ref{equx}) and $l_{\mathrm{r}}$  is given by  $l_{\mathrm{r}} =  \, r \, \frac{\omega_{\mathrm{A}}}{N}$, which is obtained by assuming  the marginal stability in   Eq.~(\ref{lr}).

 Turning towards the transport of angular momentum by  magnetic field, we first write
  the azimuthal stress  by volume unity due to the magnetic field  
  
  \begin{eqnarray}
  S \; = \frac{1}{4 \; \pi} \; B_{\mathrm{r}} B_{\varphi} \; = \;
  \frac{1}{4 \; \pi} \;  \left(\frac{l_{\mathrm{r}}}
  {r}\right) B_{\varphi}^2 = 
  \; \rho \; r^2 \; \left(\frac{\omega_{\mathrm{A}}^3}{N}\right)  \; .
  \end{eqnarray}

 \noindent
 Then, the viscosity $\nu$
 for the vertical transport of angular momentum can be expressed in terms of  $S$
 (Spruit \cite{Spruit02}),
 
 \begin{equation}
 \nu = \frac{S}{\rho \; q \; \Omega} =
  \; \frac{\Omega \; r^2}{q} \;
 \left( \frac{\omega_{\mathrm{A}}}{\Omega}\right)^3 \; 
\left(\frac{\Omega}{N}\right) \; .
 \label{nu}
 \end{equation}
 
 \noindent
 This is the general expression of $\nu$ with $\omega_{\mathrm{A}}$ given by the solution
 of Eq.~(\ref{equx}) and with $N$ by Eq.~(\ref{Nfinal}).   
 We have the full set of expressions necessary  to obtain  
 the Alfv\'en frequency $\omega_{\mathrm{A}}$ and  the magnetic diffusivity  $\eta$. The parameter
 $\eta$   also
 expresses  the vertical transport of the chemical elements, while the viscosity $\nu$
  determines the vertical transport   of the angular momentum by the magnetic field. 
  
 In paper II, we also checked that  
  the  rate of magnetic energy production 
  $W_{\mathrm{B}}=\frac{B^2}{8 \pi} \frac{\omega_{\mathrm{A}}^2}{\Omega}=
\frac{1}{2} \rho r^2 \frac{\omega_{\mathrm{A}}^4}{\Omega} $ per unit of time and volume is equal to the rate
  $W_{\nu}= {1\over 2}\rho\nu \Omega^2q^2$ of the dissipation of rotational energy by the magnetic viscosity
  $\nu$ as given above.

\section{Numerical applications} 

\subsection{The  models}

\begin{figure}[t]
  \resizebox{\hsize}{!}{\includegraphics{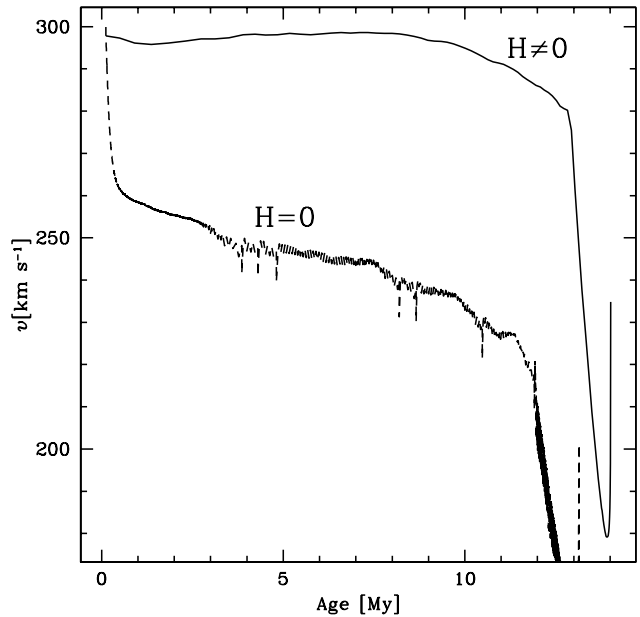}}
  \caption{Evolution of the rotation velocities at the surface of 15 M$_\odot$ models 
  during the MS phase with and without magnetic fields  $H$. The initial velocity $\upsilon_{\rm ini}$ = 300 km s$^{-1}$ in both cases. We see the much higher surface rotation when magnetic field is included.}
\label{vrot}
\end{figure}

We consider here models of 15 M$_{\odot}$, with a standard composition of $X=0.705$ and $Z=0.02$.
The physics of the models, opacities, nuclear reactions, mass loss rates, structural rotational effects,
shear mixing, meridional circulation,  etc.
are the same as in recent models of the Geneva group (Meynet \& Maeder \cite{MMXI}). 
We calculate 3 sets of  models: the first is without rotation, the second  with rotation and the third one with both rotation and magnetic field. The initial rotation velocity is 300 km s$^{-1}$, which leads to an average velocity on the MS phase of about 240 km s$^{-1}$ when no magnetic field is present (see Fig.~\ref{vrot}). 
In paper II, we did not include the effects of  the meridional currents, in order to  independently study the effects of the magnetic field. However, we showed that meridional
circulation and the magnetic field may significantly interact, with possible consequences for the transport mechanisms. Thus,  to examine this interplay, we also account here for meridional circulation currents. For the moment, we do not account for magnetic coupling by stellar winds, since there is no external
convective zone in massive stars on the Main Sequence.

The choice of the time step is 
imposed by the most rapid process taking place. We    checked that changing the time steps did not change the results.
Here the fastest  process is the  transport of the angular momentum by the magnetic diffusivity.
The diffusion coefficients $\nu$  for the angular momentum  by the magnetic field reach in some  cases values of $10^{13}$ cm$^2$ g$^{-1}$, while the average value is 1--2 orders of magnitude lower. 
The cases of strong coefficients lead to diffusion timescales $\tau \approx (\Delta R)^2/\nu \approx$ a few $10^2$ yr, where $D$ is the diffusion coefficient. For an appropriate  treatment, we need to adopt very
small time steps. In practice,  we  take  time steps of the order of 20 years. This implies 
about $6 \cdot 10^5$ models to cover in a  detailed way the exact interplay of the effects of the magnetic field and of the meridional circulation during  the MS phase !

 Clearly, the model properties do not change
significantly over such a time scale as shown for example by Fig. \ref{profoH}. Thus, in future, faster 
processes of calculations may be devised. However, by imposing very short time steps on the two processes,
we set the study in the linear regime for both instabilities  and we may thus proceed to an addition of their
own different effects.

\subsection{Evolution of the internal and superficial rotation}

 In a rotating star, the internal  profile of angular velocity $\Omega(M_{\mathrm{r}})$ changes with time during evolution, due to 
 various effects (Meynet \& Maeder \cite{MMV}): central contraction and envelope expansion, transport
 of angular momentum by circulation and mass loss at the stellar surface. One assumes here that the mass
 lost by stellar winds  just embarks its own angular momentum. The case of anisotropic stellar winds 
 has been studied by Maeder  (\cite{MIX}), who has shown that such effects are important only for 
 very massive stars ($M \geq 60$ M$_{\odot}$) in fast rotation. Here, the loss of angular momentum
  at the stellar surface only has a limited importance for the evolution of  rotation.
 Fig. \ref{profoSH} shows 
 the evolution of the internal profile of rotation. Differential rotation builds up during the MS phase, reaching about a factor of two between $\Omega$ at the surface and in the convective core, near  the end of the MS when the central H--content is $X_{\mathrm{c}}=0.10$. Then, fast central contraction accelerates the core rotation.

\begin{figure}[tb]
  \resizebox{\hsize}{!}{\includegraphics{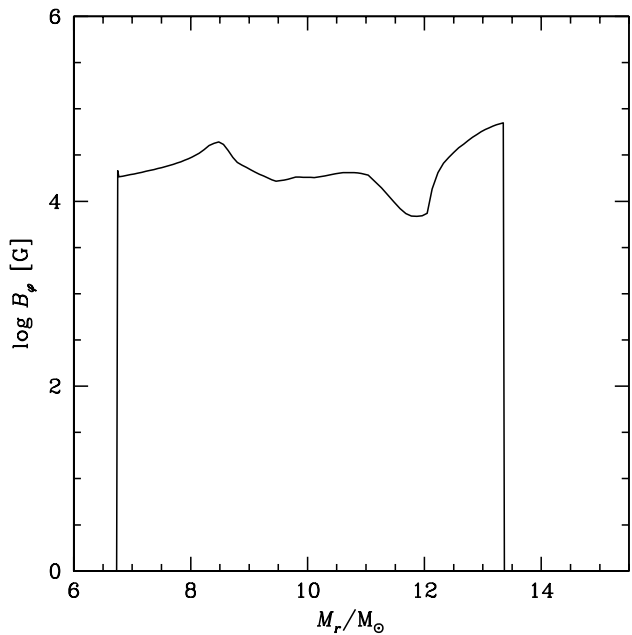}}
  \caption{Distribution of the magnetic field in the model of initial 15 M$_{\odot}$
  when the central hydrogen content is $X_{\mathrm{c}}=0.60$. The age of th model is
  $4.2798 \cdot 10^6$ yr and the actual mass is 14.97 M$_{\odot}$. }
\label{champx60}
\end{figure} 
 
 The situation is quite different when magnetic fields as described above are accounted for. As
 shown by Fig. \ref{profoH},
 the angular velocity $\Omega$ is almost constant throughout the stellar interior. 
 It is not exactly constant, otherwise
  $q$ would be zero and  the magnetic field would not be sustained anymore. Only at the very end of the MS phase does    the fast central contraction bring about a small significant difference.
  
  Fig. \ref{vrot}  shows the evolution of the surface rotation velocities as a  function of age in the 
  models with and without magnetic field. We notice the higher velocity during the whole MS phase  of the model with magnetic field, which has $v \approx 300$ km s$^{-1}$,
   compared to the model without a field, where  $v \approx 250$ km s$^{-1}$. There are 3 effects 
   intervening:
   \begin{enumerate}
\item   The main part of the
    difference already occurs  on the ZAMS. With rotation only, the model adjusts its rotation very quickly to an equilibrium profile  determined by meridional circulation (Meynet \& Maeder \cite{MMV}) and this establishes an $\Omega$--gradient which reduces the surface velocity. On the contrary, with a magnetic field,
    the model can keep a constant $\Omega$ and there is no initial decrease. 
    \item   The other two smaller  effects occur during MS evolution.  The magnetic coupling forces the surface to co--rotate with the convective core, as contraction makes it  spin--up. With rotation only,
    the coupling (due to meridional circulation) is much weaker. 
    \item In the magnetic model, the surface enrichment in helium is slightly higher, this keeps
    the radius smaller. Thus, as the radius expansion is smaller, the decrease of surface rotation is also smaller.
    \end{enumerate}

The question of the differential rotation in the horizontal direction has also to be examined.
In current rotating models without a magnetic field, it is assumed that
the horizontal turbulence is strong enough to suppress the horizontal differential rotation,
 so that the hypothesis of shellular rotation with $\Omega= \Omega(r)$ applies 
 (Zahn \cite{Zahn92}). 
In the presence of a magnetic field, the horizontal turbulence is likely reduced.
The numerical values for the field $B_{\varphi}$
 found below   is of the order of  a few $10^4$ G  (cf. Fig. \ref{champx60}). The
 horizontal  coupling ensured by the magnetic field (cf. Fig. \ref{champx60}) is expressed by 
 a coefficient $D_{\mathrm{B}_{\mathrm{h}}}$ 
(Maeder \& Meynet \cite{Magn1})

  \begin{equation}
  D_{\mathrm{B}_{\mathrm{h}}} = \;r^2 \; (\omega_{\mathrm{A}}^2/\Omega) =
   \frac{ B_{\varphi}^2}{ 4 \pi \;\rho \; \Omega}   \quad \quad \mathrm{for} \quad
   \Omega \gg \omega_{\mathrm{A}} \; .
   \label{DBhfort}
  \end{equation}
  
  \noindent
 The  value of $D_{\mathrm{B}_{\mathrm{h}}}$ is
 of the order of $10^{11}$  cm$^2/$s in the present 15 $M_{\odot}$ model.
  Thus, 
the magnetic  field is large enough to ensure the horizontal coupling, so that the
assumption of shellular rotation is still  valid.

\begin{figure}[tb]
  \resizebox{\hsize}{!}{\includegraphics{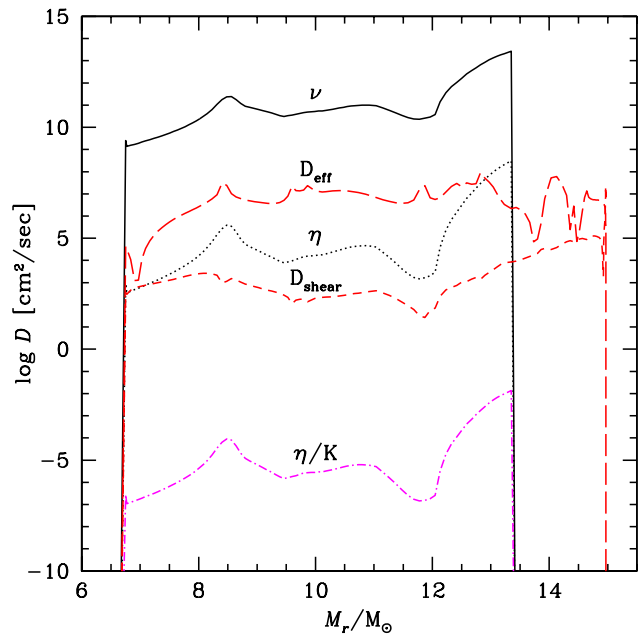}}
  \caption{The diffusion coefficients in the model with central H--content 
  $X_{\mathrm{c}}=0.60$. In the outer envelope, where the magnetic field is absent,
  the transports by shear and by combination of the meridional circulation
  and horizontal turbulence expressed by $D_{\mathrm{eff}}$ are present.}
\label{diff60}
\end{figure}

\subsection{Diffusion coefficients for the transports of the angular momentum
and  chemical elements}

We need to determine in which regions of the star the magnetic field is present, its intensity
and the run of the various diffusion coefficients.  As an example, we examine the
15 M$_{\odot}$ model at the beginning of the evolution, when the central H--content is
$X_{\mathrm{c}}=0.60$. In this model as in all MS models,  $q \, \geq \, q_{\mathrm{min}}$ in a region 
which starts just above the core and extends over most of the envelope. In the model with $X_{\mathrm{c}}=0.60$, the magnetic field is present  from
$M_{\mathrm{r}} = 6.7$ M$_{\odot}$ to 13.4 M$_{\odot}$, 
as illustrated in Fig. \ref{champx60}.   The average field is about 
$2 \cdot 10^4$ G and the corresponding value of $\omega_{\mathrm{A}}/\Omega$ is about
$5 \cdot 10^{-4}$. Above 13.4 M$_{\odot}$, condition (\ref{cond1}) is no longer
satisfied. The region where magnetic field is present slightly increases during MS evolution
as the convective core recedes. Near the end of the MS phase, when $X_{\mathrm{c}}=0.05$,
the magnetic field is present from 4.2 M$_{\odot}$ up to 13.8 M$_{\odot}$.

 \begin{figure}[tb]
  \resizebox{\hsize}{!}{\includegraphics{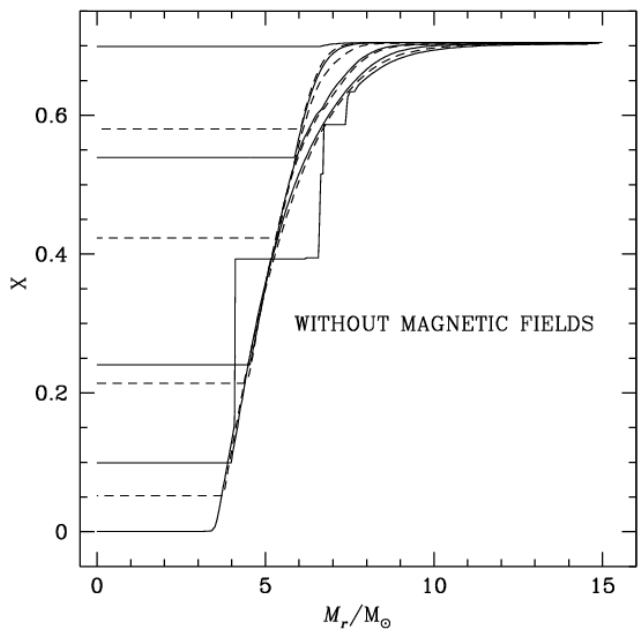}}
  \caption{ Internal distribution of the hydrogen mass fraction $X$
   as a function of the Lagrangian mass in a 15 M$_\odot$ model  with rotation
    at various stages of the model evolution from top to bottom during the MS--phase.
    Some convective zones appear in the H--shell burning stage at the end of the MS phase.}
\label{profxSH}
\end{figure}

\begin{figure}[tb]
  \resizebox{\hsize}{!}{\includegraphics{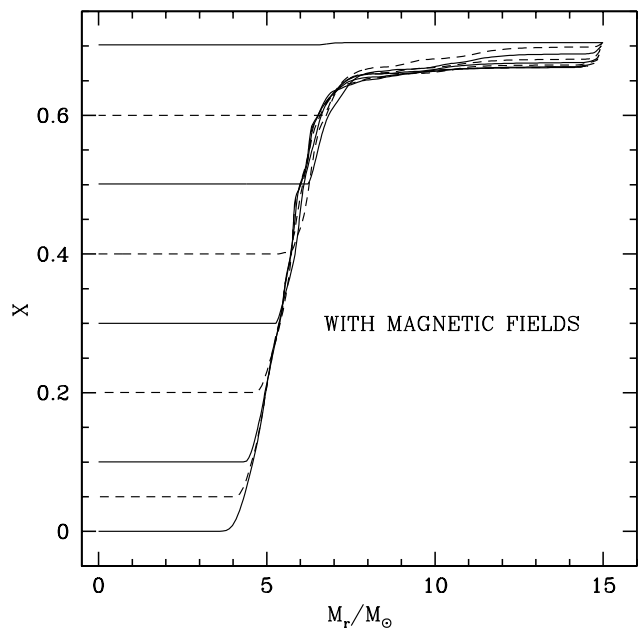}}
  \caption{Internal distribution of the hydrogen mass fraction $X$
  as a function of the Lagrangian mass in the model   with magnetic field calculated according to
  expressions of the present paper at various stages of the model 
  evolution from top to bottom during the MS--phase.}
\label{profxH}
\end{figure}

The runs of the various diffusion coefficients are illustrated in Fig. \ref{diff60}. The largest diffusion  coefficient
is $\nu$ which acts for the vertical transport of angular momentum, the large value of $\nu$
 imposes the  nearly constant $\Omega$ in the interior. The value  of $\nu$ is about 6 orders of magnitude larger  than the diffusion coefficient $\eta$ for the transport of the chemical elements, so that the surface enrichments in products of the CNO cycle due to this magnetic diffusion coefficient only are rather limited. The coefficient 
$D_{\mathrm{eff}}$ which applies to the transport of chemical elements by meridional circulation is much larger than $D_{\mathrm{shear}}$, while in rotating stars without magnetic field, it is generally 
the opposite. This is due to the fact that the velocity of meridional circulation is much larger when $\Omega$ is almost constant throughout the stellar interior (about $10^{-2}$ cm s$^{-1}$),
while  $D_{\mathrm{shear}}$ is much smaller.
Indeed $D_{\mathrm{shear}}$  is about 4  orders of magnitude smaller than in rotating stars
 without a magnetic field, as  a consequence of the near solid body rotation in magnetic models.
  
 Fig. \ref{diff60} also shows that the value $\eta/K$ is always very small, which justifies the simplifications made in deriving Eq. (\ref{equx}). Apart from the slight extension of the zone covered by the magnetic field as mentioned above, the orders of magnitude and relative ratios of the diffusion coefficients remain about the same during the whole MS evolution.
 
An important result of Fig. \ref{diff60} is that the transport  by meridional circulation (expressed by
$D_{\mathrm{eff}}$) is in general 2--3 orders of magnitude larger than the transport of the elements 
 by the magnetic field (expressed by $\eta$). This clearly
shows that the direct effect of the magnetic instability is of little importance with respect to thermal instability, which 
drives meridional circulation, for the transport of chemical elements. In addition, meridional circulation  is important for the transport of the elements from the external edge of the magnetic zone to the stellar surface.
However, for the coupling of $\Omega$, the magnetic field is much more efficient than meridional circulation.
One can define a velocity $U_{\mathrm{B}}$ for the vertical transport of angular momentum  by the magnetic field (cf. paper I)

\begin{equation}
U_{\mathrm{B}}= 5 \; \frac{\nu}{r} \; \frac{\partial \ln \Omega}{\partial \ln r}  \; .
\label{UB}
\end{equation}

\noindent
This quantity may be compared to the vertical component of the
velocity $U(r)$ of meridional circulation. In general  $U_{\mathrm{B}}$ 
is much larger than $U(r)$, which confirms the dominant role of the magnetic field for the transport of angular momentum.

 \subsection{Internal distribution of hydrogen}

\begin{figure}[tb]
  \resizebox{\hsize}{!}{\includegraphics[angle=0]{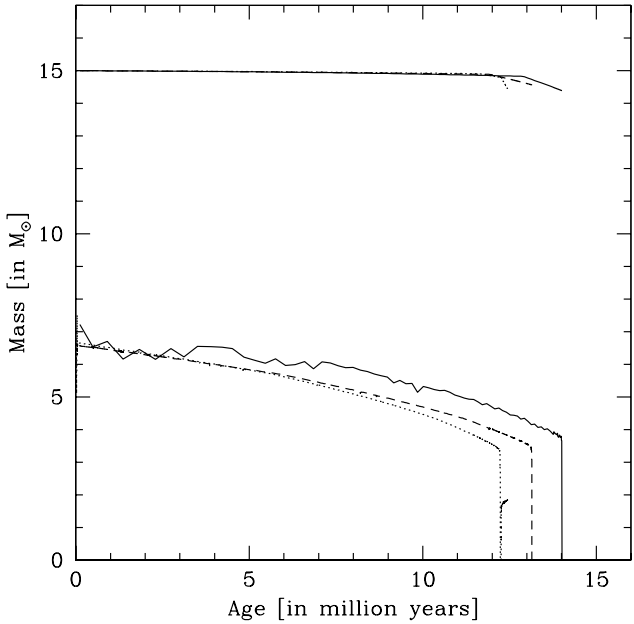}}
  \caption{Top lines near 15 M$_{\odot}$: evolution of the stellar mass. Lower lines: 
  evolution of the mass of the convective core. The dotted line
  applies to the  model without rotation, the short--broken line to the model with rotation 
  ($\upsilon_{\rm ini}$= 300 km s$^{-1}$) but without
  magnetic fields, the continuous 
  line to the model with rotation ($\upsilon_{\rm ini}$= 300 km s$^{-1}$) and magnetic fields.}
\label{kip}
\end{figure}

Figs. \ref{profxSH} and \ref{profxH} show the internal distribution of hydrogen in the cases without and with magnetic field.   
Firstly, rotation with a magnetic field 
leads to the formation of a larger core at the end of the MS phase compared to models with rotation only
(models with rotation  themselves  have larger cores compared to models without rotation, cf. also
Fig. \ref{kip}).

In models with rotation only, there is a significant erosion of the $\mu$--gradient at the 
immediate edge of the core, developing during the second half of
the MS phase. This erosion directly at the edge of the core is absent in models with a magnetic field.
The reason is Eq. (\ref{eta}), which gives, if the $\mu$--term dominates in the
Brunt--V\H{a}is\H{a}l\H{a} frequency ,

\begin{equation}
\eta_0 = r^2 \Omega  q^4 \; \left(\frac{\Omega}{N_{\mu}}\right)^6  \; .
\label{etaz0}
\end{equation}

\noindent
As noted in paper II, the mixing of chemical elements decreases
strongly if the $\mu$--gradient grows and this effect  limits
the chemical mixing of elements by the Tayler--Spruit dynamo in the regions
just above the convective core.

\begin{figure}[t]
  \resizebox{\hsize}{!}{\includegraphics[angle=0]{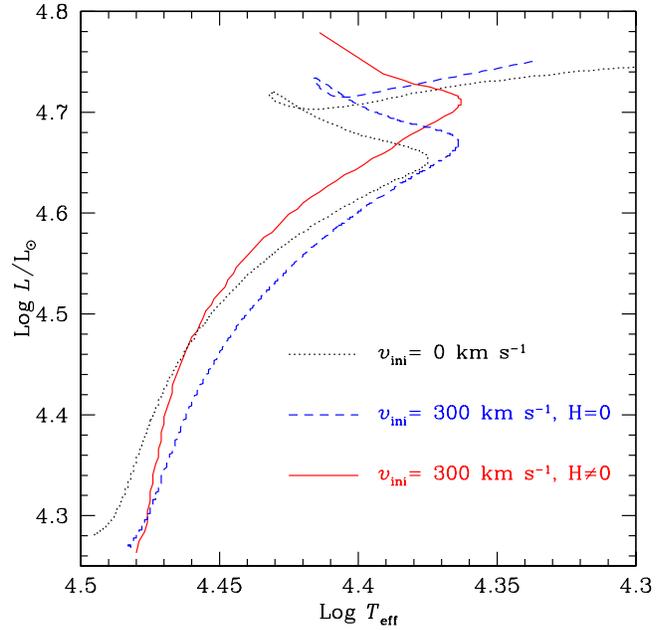}}
  \caption{ Evolutionary tracks in the HR diagram for 15 M$_\odot$ stellar models. The dotted line
  applies to the  model without rotation, the short--broken line to the  model with rotation 
  ($\upsilon_{\rm ini}$= 300 km s$^{-1}$) but without
  magnetic fields, the continuous 
  line to the model with rotation ($\upsilon_{\rm ini}$= 300 km s$^{-1}$) and magnetic fields.  }
\label{dhr}
\end{figure}

\begin{figure*}[t]
  \resizebox{\hsize}{!}{\includegraphics[angle=-90]{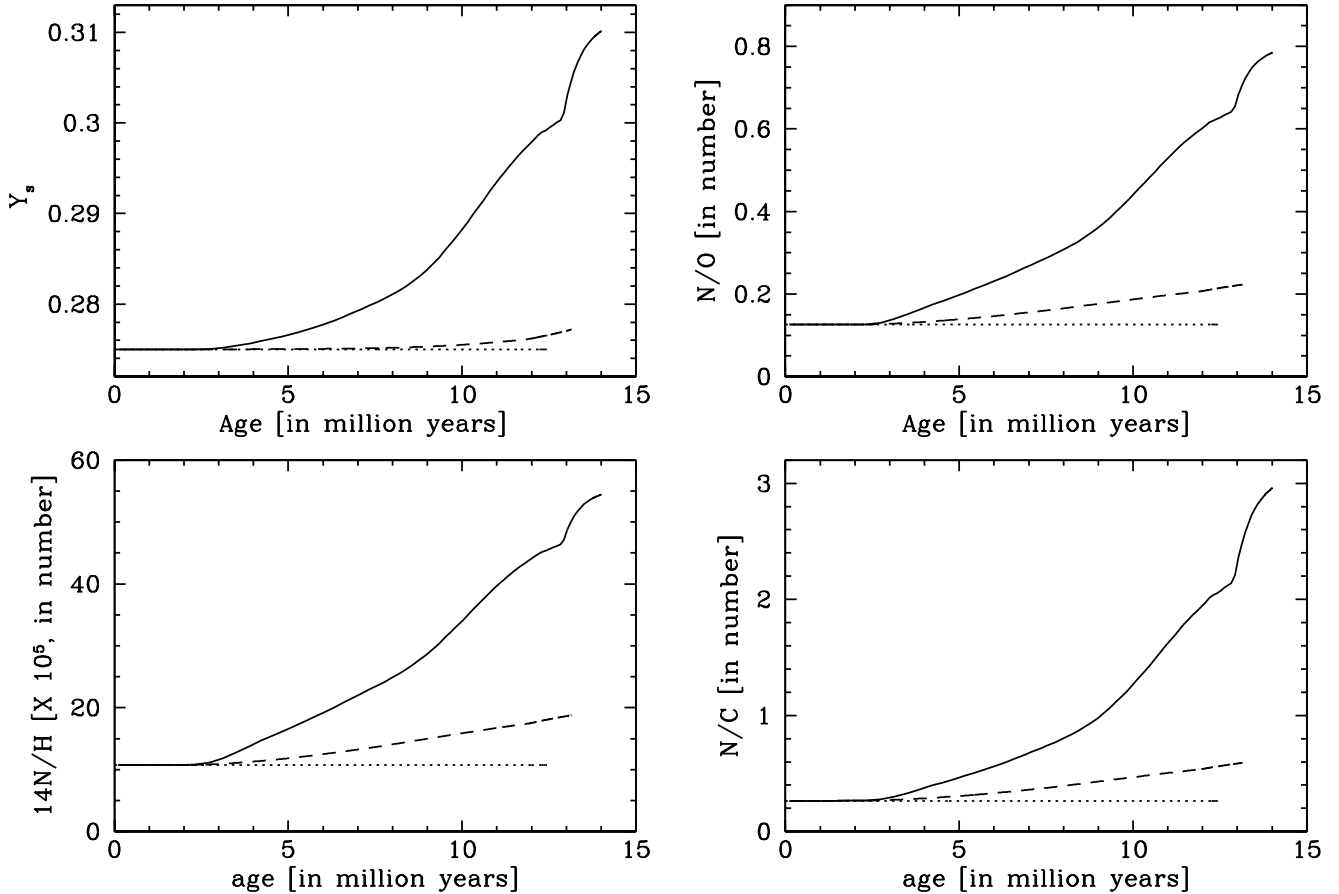}}
  \caption{Time evolution of the surface helium content $Y_{\mathrm{s}}$ in mass fraction, of the 
  N/O, N/H and N/C in mass fraction for various models: The dotted line
  applies to the  model without rotation, the short--broken line to the model with rotation 
  ($\upsilon_{\rm ini}$= 300 km s$^{-1}$) but without
  magnetic fields, the continuous 
  line to the model with rotation ($\upsilon_{\rm ini}$= 300 km s$^{-1}$) and magnetic fields.  }
 \label{abond}
\end{figure*}

Finally, in models with a magnetic field the mixing in the envelope is much greater than in 
the case without a field.
This effect appears  during the entire MS phase and the final helium surface content reaches 
$Y_{\mathrm{s}} = 0.31$. This effect is not due to the shear, which is essentially absent
in magnetic models. It is also not  due to the magnetic diffusivity, because it is rather small and
in the very external regions there is no
magnetic field. However, the transport of chemical elements by the circulation
(an effect described by $D_ {\mathrm{eff}}$) is  larger and is generally the main effect in the envelope. It
produces a slight transport of the elements, which enriches the stellar surface in elements
of the CNO cycle (cf. Fig. \ref{abond}). The models of paper II with a magnetic field,
 but without meridional circulation  predict no N excesses and this was a difficulty.
However, magnetic models, when the meridional circulation is included, lead
to a significant N--enrichment, which  compares  with observations (cf. Sect.
\ref{chimab}).

 \section{Evolution of observable parameters}
 
 The magnetic field created by the dynamo does not reach the stellar surface. However,
 there are consequences concerning the tracks in the HR diagram and especially the 
 enrichment of surface abundances, which may lead to observable consequences.
 
\subsection{Evolution of the core, lifetimes and  tracks in the HR diagram}

The evolution of the tracks in the HR diagram is determined by the evolution of the core 
mass fraction. Fig. \ref{kip} shows the evolution of the total stellar mass in the magnetic case. 
When $X_{\mathrm{c}}=0.30$, the actual mass is 14.875 M$_{\odot}$. Mass loss increases near the end of the MS phase and the mass  is 14.389 M$_{\odot}$ when $X_{\mathrm{c}}=0.000$. As is well known, rotational effects
slightly increase the core mass fraction near the end of the MS phase and produces longer 
MS lifetimes. The account of the magnetic field further enlarges the core mass fraction
as well as the MS lifetimes as illustrated in Fig. \ref{kip}.

The effects on the tracks in the HR diagram are shown in Fig. \ref{dhr}. In addition to the structural changes,
rotation also produces a distortion of the stellar surface and an increase of the average radius (estimated at
$P_2 (\cos \vartheta)=0$, where $\vartheta$ is the colatitude). This makes a redwards shift  
of the tracks of rotating models  in the HR diagram, especially visible near the ZAMS, since  no
other effects intervene there.  For the model with rotation only, we see  the upwards 
 shift due to the slightly larger core 
and a slightly bluer track resulting from a tiny surface He--enrichment.
The model with rotation and magnetic field reaches a higher luminosity at the end of the MS due to the
larger core. It is also bluer due to the larger surface enrichment in helium, which lowers the opacity and decreases the stellar radius. 
It is doubtful that the above effects are sufficient  to infer the presence or absence of
 magnetic fields.

\subsection{Evolution of the abundances of helium and CNO 
elements at the stellar surface}   \label{chimab}

The evolution of the surface abundances in helium and CNO elements results from the internal profiles
shown in Figs. \ref{profxSH} and \ref{profxH}. Fig. \ref{abond} shows for the 3 models considered
the evolution of the helium abundance $Y_{\mathrm{s}}$ in mass fraction at the surface, the $N/H$, $N/O$ and $N/C$ ratios of numbers of atoms. We see the same trends in the four panels. There is no enrichment
in absence of rotation. With rotation only, there are moderate enrichments, by a factor of 2
for $N/H$,  2.5 for $N/C$ and 1.8 in $N/O$. With both rotation and magnetic field, the
surface enrichments are much larger, particularly for the helium abundance which reaches 
$Y_{\mathrm{s}}=0.31$ at the end of the MS phase. The increases in $N/H$ reach a factor of 5, 
11 for $N/C$ and 6 for $N/O$.

Observationally, there are  many estimates of N enrichments for OB stars. A recent 
review of the subject has been made by Herrero \& Dufton (\cite{HerrDuft04}), who show clear evidence
of rotationally--induced mixing in OB stars,  also with the result that some fast rotating 
stars do not show N--enrichments.
Four O9 stars were studied by Villamariz et al.  (\cite{Villamariz02}); three low rotators have 
an excess of $[N/O] \, < \, 0.2$ dex, one fast rotating O9III star with $v \sin i =430$
km s$^{-1}$ has an excess of $[N/O] \, = \, 0.7$ dex. The two most massive stars with fast rotation in 
the association Cep 2 have an excess of $[N/O] \, < \, 0.3$ dex ( Daflon et al.  
\cite{Daflon01}). Two stars in Sher 25 have excesses in $[N/O]$ by  a factor of 3 to 4 
(Smartt et al. \cite{Smartt02}). Observations of the $N/H$ ratios by Venn and 
Przybilla (\cite{VennPr03}) for galactic A--F supergiants show an average excess of a factor of 3, with extreme values up to a factor of 8.
The orders of magnitude of the predicted and observed enrichments 
are  similar. However, the situation is still uncertain due to the relative lack of accurate observational data for MS stars.

 \section{Conclusions}
 
 The main result is that a magnetic field imposes nearly solid body rotation and this favours higher
 rotational velocities during MS evolution compared to cases where the  magnetic field is not accounted for.
 Due to the nearly constant $\Omega$ in the stellar interior, the transport of  chemical elements by 
 shear mixing is negligible. The transport of  elements by Tayler--instability is 
 also very limited. An interesting feature of the model is that meridional circulation is 
 strongly enhanced by the flat $\Omega$--curve and this is the main effect influencing  the transport of the chemical
 elements in the present models.
 
 There remain however some doubts as to whether the Tayler--Spruit dynamo is really  active 
 in  stellar interiors. Up to now, magnetohydrodynamic models expressing the growth and evolution of the 
 magnetic field in  rotating stars have not yet confirmed the existence and efficiency of this
 particular instability in stellar interiors (Mathis \cite{Mathis05}). Even  in the observationally and theoretically much better studied case of the Sun,
 the exact location, origin and evolution of the solar dynamo are still not fully understood.
 
 The larger size of the predicted surface enrichments 
 is well (or even
 better)  supported by observations. However, this remains uncertain in view of the small number of accurate 
 observations.  In this respect, it might be crucial to also have such comparisons for stars in the Magellanic Clouds where the observed  enrichments are much larger than in the Galaxy.
The answer may come from asteroseismology. There are no p--modes expected, however g--modes may be 
present and yield some information on the internal $\Omega$--distribution. At present, this seems to be 
the most compelling possible test.



\appendix
\section{Equations for the case of  low rotation}

 The dynamo properties were established (Spruit \cite{Spruit02} and Sect. 2) for the case of fast rotation, with
  the condition that the rotation rate $\Omega$ is larger than the 
 Alfv\'en frequency $\omega_{\mathrm{A}}$, i.e.  
$ \Omega \gg \omega_{\mathrm{A}}$.
In the fast rotating  case, the growth rate of the magnetic field is reduced by a factor 
$\omega_{\mathrm{A}}/\Omega $ as firstly suggested by Pitts \& Tayler (\cite{Pitts86}).
The above equations of the dynamo have been derived under this condition. 
 However, if we want to study very slowly rotating stars, we need also
to consider the case where 

\begin{equation}
 \Omega  \ll \omega_{\mathrm{A}}
 \label{regime2}
 \end{equation}
 
  \noindent
This may be interesting also in central regions of radiative stars. 
 There, as a result of the small value of $r$, 
 the Alfv\'en frequency $\omega_{\mathrm{A}} $ becomes large so that the regime corresponding 
 to Eq.~(\ref{regime2}) may apply instead of the usual one (cf. Sect. \ref{dyn}). 

The energy of  the Tayler instability
(Tayler \cite{Tay73}) must be large enough to overcome the restoring force of 
buoyancy and this implies  that the Alfv\'en frequency must be larger than some limit
depending on $N$.
The vertical extent $l_{\mathrm{r}}$ of the magnetic instability is also given by Eq. (\ref{lr}) above.
In order that the perturbation is not too quickly damped 
 by the diffusion of the magnetic field, one must have 
$ l_{\mathrm{r}}^2 >  \frac{\eta}{\sigma_{\mathrm{B}}}$. At low rotation,
 in the  absence of significant Coriolis force, the frequency $\sigma_{\mathrm{B}}=\omega_{\mathrm{A}}$. 
 The combination of these two limits in the low rotation case leads to the 
condition  $\omega_{\mathrm{A}}^3 >  N^2\;
\frac{\eta}{r^2}$ and  for  the  marginal situation corresponding to equality, we have

\begin{equation}
\left(\frac{\omega_{\mathrm{A}}}{\Omega}\right)^3 =  \frac{N^2}{\Omega^2} \;
\frac{\eta}{r^2 \; \Omega}   \; .
\label{premierA}
\end{equation} 

 \noindent  This equation relates  the magnetic diffusivity 
 $\eta$ and the Alfv\'en frequency $\omega_{\mathrm{A}}$ instead of Eq. (\ref{premier}).
 
 The field amplitude may be fixed by the equality of the  amplification 
 time $\tau_{\mathrm{a}}$ of the Tayler  instability and of the  timescale
 $\sigma_{\mathrm{B}}^{-1}$  of 
 the field (Spruit \cite{Spruit02}). One has
 
 \begin{equation}
 \tau_{\mathrm{a}}=  N /(\omega_{\mathrm{A}} \Omega q)
 \end{equation}
 
 \noindent
 Instead of the expression in (\ref{qo}), 
 we get for the low rotation case with $\sigma_{\mathrm{B}}=\omega_{\mathrm{A}}$

\begin{equation}
\frac{N}{\Omega} \; = \; q   \; .
\label{Nq}
\end{equation}

\noindent
$\omega_{\mathrm{A}}$ has disappeared from the equation.
How should we interpret this ? This can be done with the help of Eq.(\ref{Nfinal}), which gives 

\begin{equation}
\frac{ N^2_{\mathrm{T}}}{\Omega^2} \; \frac{\eta / K} {\eta / K  \;+ \; 2} + 
\frac{N^2_{\mu}}{\Omega^2} =q^2.
\label{NNq}
\end{equation} 

\noindent
The two equations (\ref{premierA}) and  (\ref{NNq}) form our basic system  of equations with 
the two unknown quantities  $\eta$ and $\omega_{\mathrm{A}}$. In order to have a diffusivity $\eta$  positively defined, Eq.~(\ref{NNq})
 imposes that the differential rotation parameter $q$ must
 be larger than some minimum value $q_{\mathrm{min}}$, 

\begin{equation}
q^2_{\mathrm{min}}\;  = \frac{N^2_{\mu}}{\Omega^2}  \; .
\label{qmin}
\end{equation}

\noindent
If this is the case, then  according
to Eq.~(\ref{premierA}) the same is true for $\omega_{\mathrm{A}}$ and thus 
a magnetic field is effectively  created by the slow  dynamo process. 

Let us now estimate the transport coefficients.
Eq.~(\ref{NNq}) immediately leads to the following expression for  the magnetic diffusivity $\eta$,

\begin{equation}
\eta = \frac{ 2 \; K 
\left( q^2 \; \Omega^2 - N^2_{\mu}\right)}
{N^2_{\mathrm{T}}+N^2_{\mu} - q^2 \; \Omega^2} \; .
\label{etageneralA}
\end{equation}

\noindent
In all  cases, the ratio $\eta/K$ is very small, typically of the order of $10^{-5}$ (cf. Fig. \ref{diff60}). Thus, 
we obtain  the corresponding expression for the diffusivity,

\begin{equation}
\eta = \frac{2 \; K}{N^2_{\mathrm{T}}} \; 
\left( q^2 \; \Omega^2 - N^2_{\mu}\right) \; .
\label{etautilA}
\end{equation}

\noindent
 This equation  provides the
 magnetic diffusivity  $\eta$,  if $|q|$
 is larger than $|q_{\mathrm{min}}|$. The coefficient $\eta$  
 determines the transport of the chemical elements by
 the magnetic instability.
 
   Now, with Eq.~(\ref{premierA}), we   obtain  
 the Alfv\'en frequency $\omega_{\mathrm{A}}$
at each location $r$ in the star,

\begin{equation}
\left(\frac{\omega_{\mathrm{A}}}{\Omega}\right)^3= \frac{2 \; K}{N^2_{\mathrm{T}}} \;
\frac   {\left(q^2 \; \Omega^2 - N^2_{\mu}\right)}   {r^2 \; \Omega}  \; q^2 \; .
\label{omegaAfinal}
\end{equation}

\noindent
This confirms that for a star rotating with angular velocity $\Omega$ and
having a certain $\mu$--gradient, a magnetic field is created only 
if the actual value of the differential rotation parameter $|q|$ is larger than
$|q_{\mathrm{min}}|$. 

 The azimuthal stress $S$ due to the magnetic field generated by Tayler instability becomes

 \begin{eqnarray}
  S \; = \frac{1}{4 \; \pi} \; B_{\mathrm{r}} B_{\varphi} \; = \;
  \frac{1}{4 \; \pi} \;  \left(\frac{l_{\mathrm{r}}}
  {r}\right) B_{\varphi}^2 = 
  \; \rho \; r^2 \; \left(\frac{\omega_{\mathrm{A}}^3}{N}\right)  \; .
  \end{eqnarray}
  
 \noindent
  The viscosity $\nu$ is expressed in terms of  $S$ 
and we get 
 
  \begin{equation}
 \nu =  \frac{S}{\rho \; q \; \Omega}\,  = \,\frac{\Omega \; r^2}{q} \;
 \left( \frac{\omega_{\mathrm{A}}}{\Omega}\right)^3 \; 
\left(\frac{\Omega}{N}\right) \; .
\label{nugeneralA}
\end{equation} 

\noindent
With   Eq.~(\ref{omegaAfinal}) for the Alfv\'en frequency,
 we get finally
\begin{equation}
\nu =  \frac{\Omega \; r^2}{q^2} \;
 \left( \frac{\omega_{\mathrm{A}}}{\Omega}\right)^3 \; =
 \frac{2 \; K}{N^2_{\mathrm{T}}} \;
  {\left(q^2 \; \Omega^2 - N^2_{\mu}\right)}    \; ,
\label{nufinalA}
\end{equation}
\noindent
which determines  the transport of angular momentum.
The expression of $\nu$ is the same as the
expression of $\eta$ (Eq.~\ref{etautilA}). This means that in the  case of
low rotation, the Tayler magnetic instability transports in a similar way the chemical elements
and the angular momentum, while at  high rotation the magnetic coupling is  stronger
by a factor $\left(\frac{\Omega}{\omega_{\mathrm{A}}}\right)^2$
than the transport of the elements, according to the general expressions by Maeder \& Meynet
(\cite{Magn2}). This factor comes from the expression of $\sigma_{\mathrm{B}}$,
and as in the present case the ratio  $\left(\frac{\Omega}{\omega_{\mathrm{A}}}\right)$
is absent in $\sigma_{\mathrm{B}}$, we get equality of $\nu$ and $\eta$.

 Condition (\ref{Nq}) must also be satisfied in order to have angular momentum transport.
The equations (\ref{etageneralA}) or (\ref{etautilA}) and ({\ref{nufinalA}) provide the transport coeffcients
at each stellar layer as a function of the 
local  quantities, such as $\Omega$, $q$, $N^2_{\mathrm{T}}$,
$N^2_{\mu}$, etc... available in the stellar models. The Alfv\'en frequency
and the magnetic field intensity are obtained by Eqs. (\ref{omegaAfinal})   and (\ref{def1}).

The rate of magnetic energy production 
  $W_{\mathrm{B}}$ per unit of time and volume must be equal to the rate
  $W_{\nu}$ of the dissipation of rotational energy by the magnetic viscosity
  $\nu$ as given above. This check of consistency was verified for the case of high rotation.
  One assumes here for simplification  that all the  energy 
  dissipated is converted to magnetic energy. One has
  
  \begin{equation}
W_\nu={1\over 2}\rho\nu \Omega^2q^2 ,
\label{Wnu}
\end{equation}

\noindent
which  with Eq.~(\ref{nufinalA}) gives the dissipation rate,

\begin{equation}
W_\nu= \rho \; \Omega^2 \, q^2 \;  \frac{ K}{N^2_{\mathrm{T}}} \;
  {\left(q^2 \; \Omega^2 - N^2_{\mu}\right)}    \; .
\label{nunu}
\end{equation}

\noindent

The magnetic energy per unit volume is $\frac{B^2}{8 \pi}$, it
is produced in a characteristic time given by $\sigma_{\mathrm{B}}^{-1} =
 \omega_{\mathrm{A}}^{-1}$. Thus, one has

\begin{equation}
W_{\mathrm{B}}= \frac{B^2}{8 \pi} \omega_{\mathrm{A}}=
\frac{1}{2} \rho r^2 \omega_{\mathrm{A}}^3\
=  \rho  q^2  \Omega^2   \frac{ K}{N^2_{\mathrm{T}}} \;
  {\left(q^2 \; \Omega^2 - N^2_{\mu}\right)}    \; ,
\label{WB}
\end{equation}

\noindent
where we have used the Eq.~(\ref{omegaAfinal}) for the Alfv\'en frequency.
Thus,
 the expressions for $W_{\nu}$ and $W_{\mathrm{B}}$ are the same. We see that no
   magnetic energy is produced  if condition (\ref{qmin})
is not realized. This shows the consistency
of the field expression for $B_{\varphi}$ and of the transport coefficient
$\nu$.

\end{document}